\RequirePackage[OT1]{fontenc}
\documentclass[journal]{IEEEtran}
\usepackage{cite}

\ifCLASSINFOpdf
  \usepackage[pdftex]{graphicx}
\else
\fi
\usepackage{amsmath}
\usepackage{amssymb}
\usepackage{mathtools}
\usepackage{siunitx}
\usepackage{xcolor}

\DeclareMathOperator{\tr}{tr}
\DeclareMathOperator{\mse}{mse}
\DeclareMathOperator{\mmse}{mmse}

\newcommand{\DKL}[2]{D_\text{KL}\bigl(#1 \Vert #2\bigr)}

\newtheorem{lemma}{Lemma}
\newtheorem{theorem}{Theorem}
\newtheorem{corollary}{Corollary}

\begin{document}
%
\title{MMSE Bounds Under Kullback--Leibler Divergence Constraints on the Joint Input-Output Distribution}
%
%
%

\author{Michael~Fau\ss{},~\IEEEmembership{Member,~IEEE,}
        Alex~Dytso,~\IEEEmembership{Member,~IEEE,}
        and~H.~Vincent Poor,~\IEEEmembership{Fellow,~IEEE}
\thanks{M.~Fau\ss{}, A.~Dytso and H.~V.~Poor are with the Department of Electrical Engineering, Princeton University, Princeton, NY 08544, USA. E-mail: \{mfauss,adytso, poor\}@princeton.edu.}
\thanks{This work was supported in part by the U.S.~National Science Foundation under Grant CCF-1908308.}
\thanks{The work of M.~Fau{\ss} was supported by the German Research Foundation (DFG) under Grant 424522268.}
}

%
%

\markboth{Journal of \LaTeX\ Class Files,~Vol.~X, No.~X, August~202X}%
{Shell \MakeLowercase{\textit{et al.}}: Bare Demo of IEEEtran.cls for IEEE Journals}
%



\maketitle

\begin{abstract}
  This paper proposes a new family of lower and upper bounds on the minimum mean squared error (MMSE). The key idea is to minimize/maximize the MMSE subject to the constraint that the joint distribution of the input-output statistics lies in a Kullback--Leibler divergence ball centered at some Gaussian reference distribution. Both bounds are tight and are attained by Gaussian distributions whose mean is identical to that of the reference distribution and whose covariance matrix is determined by a scalar parameter that can be obtained by finding the root of a monotonic function. The upper bound corresponds to a minimax optimal estimator and provides performance guarantees under distributional uncertainty. The lower bound provides an alternative to well-known inequalities in estimation theory, such as the Cram\'{e}r--Rao bound, that is potentially tighter and defined for a larger class of distributions. Examples of applications in signal processing and information theory illustrate the usefulness of the proposed bounds in practice.
\end{abstract}


%
\IEEEpeerreviewmaketitle

\section{Introduction and Problem Formulation}
\label{sec:problem}

The mean square error (MSE) is a natural and commonly used measure for the accuracy of an estimator. The minimum  MSE (MMSE)  plays a central role in statistics \cite{Lehmann1998, Dodge2008}, information theory \cite{Guo2011, Dytso2018}, and signal processing \cite{Kay1993, Dalton2012, Dalton2012b} and has been shown to have close connections to entropy and mutual information \cite{Guo2005, Guo2006}.

However, often the MMSE is difficult to compute so that bounds have to be considered instead. Generally, MMSE lower bounds can be broken into three families. The first family, termed Ziv--Zakai bounds, works by connecting estimation and binary hypothesis testing \cite{ziv1969some}. The second family, termed Weiss--Weinstein, works by using the Cauchy-Schwartz inequality \cite{weinstein1988general}; the ubiquitous Cram\'er-Rao bound is an example of this family. Finally, the third family of lower bounds, termed the genie approach, works by providing side information and, thus, reducing the MMSE \cite{flam2012mmse}. The most popular approach to finding upper bounds on the MMSE works by choosing some sub-optimal estimator (e.g. linear) that renders the MMSE computable. Another, less common approach works by finding the least favorable distribution, which in some cases leads to computable MMSEs \cite{dytso2018structure}. 

In \cite{DytsoFauss2019}, we presented upper and lower bounds on the MMSE of additive noise channels when the input distribution is close to a Gaussian reference distribution in terms of the Kullback--Leibler (KL) divergence, also known as relative entropy. In this paper, the results in \cite{DytsoFauss2019} are generalized to arbitrary channels. This is accomplished by requiring the \emph{joint} input-output distribution to be close to a Gaussian reference distribution, again in terms of the KL divergence. The obtained bounds are shown to be tight and to be attained by jointly Gaussian distributions, whose mean is identical to that of the reference distribution and whose covariance matrix can be determined by finding the scalar root of a monotonic function. In analogy to the findings in \cite{DytsoFauss2019}, the bounds are shown to correspond to the best case (fundamental accuracy limit) and worst case (minimax robust solution) scenario among the feasible distributions. 

More formally, let $(\mathbb{R}^K, \mathcal{B}^K)$ denote the $K$-dimensional Borel space, and let $X \in (\mathbb{R}^K, \mathcal{B}^K)$ and $Y \in (\mathbb{R}^{M}, \mathcal{B}^{M})$ be two random variables with joint distribution $P$. The MSE when estimating $X$ form $Y$ is defined as a function of the joint distribution $P$ and an estimator $f$, that is,
\begin{equation}
  \label{eq:def_mse}
  \mse_{X | Y}(f, P) \coloneqq E_P\Bigl[ \left\lVert X - f(Y) \right\rVert^2\Bigr],
\end{equation}
where $E_P$ denotes the expectation taken with respect to $P$ and $f$ denotes a measurable function mapping from $(\mathbb{R}^K, \mathcal{B}^K)$ to $(\mathbb{R}^M, \mathcal{B}^M)$. The set of all estimators\footnote{More precisely, $\mathcal{F}$ denotes a quotient set, where two estimators are equivalent if they differ only on an Lebesgue null set.} is denoted by $\mathcal{F}$, and the MMSE is defined as
\begin{equation}
  \mmse_{X | Y}(P) \coloneqq \inf_{f \in \mathcal{F}} \; \mse_{X | Y}(f, P).
\end{equation}
The estimator attaining the MMSE is $f(Y) = E_P\bigl[ X | Y \bigr]$, the latter denoting the expected value of $X$ given $Y$ under $P$.

The problems investigated in this paper are
\begin{align}
  \label{eq:problem_sup}
  \sup_{P} \; \mmse_{X|Y}(P) \quad &\text{s.t.} \quad P \in \mathcal{P}_\varepsilon(P_0), \\
  \label{eq:problem_inf}
  \inf_{P} \; \mmse_{X|Y}(P) \quad &\text{s.t.} \quad P \in \mathcal{P}_\varepsilon(P_0),
\end{align}
where $\mathcal{P}_\varepsilon(P_0)$ is a KL divergence ball of radius $\varepsilon$ centered at $P_0$, that is,
\begin{equation}
  \mathcal{P}_\varepsilon(P_0) \coloneqq \bigl\{ P : \DKL{P}{P_0} \leq \varepsilon \bigr\}.
\end{equation}
Moreover, $P_0$ is assumed to be a Gaussian distribution,
\begin{equation}
  \label{eq:P0_Gaussian}
  P_0 = \mathcal{N}(\mu_0, \Sigma_0),
\end{equation}
with mean vector
\begin{equation}
  \mu_0 = \begin{bmatrix} \mu_{X_0} \\ \mu_{Y_0} \end{bmatrix} \in \mathbb{R}^{K+M},
  \label{eq:mu0}
\end{equation}
$\mu_{X_0} \in \mathbb{R}^K$, $\mu_{X_0} \in \mathbb{R}^M$, and covariance matrix
\begin{equation}
  \label{eq:Sigma0_blocks}
  \Sigma_0 = \begin{bmatrix} A_0 & B_0 \\ B_0^\text{T} & C_0 \end{bmatrix} \in \mathbb{S}^{K+M},
\end{equation}
where $A_0 \in \mathbb{S}^K$, $B_0 \in \mathbb{R}^{K \times M}$ and $C_0 \in \mathbb{S}_+^M$. Here $\mathbb{S}_+^K$ ($\mathbb{S}^K$) denotes the sets of real positive (semi)definite matrices of size $K \times K$. 

The main results of the paper, the solutions of \eqref{eq:problem_sup} and \eqref{eq:problem_inf}, are stated in Section~\ref{sec:main_result}, proved in Sec.~\ref{sec:proof} and are briefly discussed in Section~\ref{sec:discussion}. Some illustrative numerical examples are presented in Section~\ref{sec:examples}. Section~\ref{sec:conclusion} concludes the paper and discusses open problems for future research.

\section{Main Result}
\label{sec:main_result}

Before stating the solutions of \eqref{eq:problem_sup} and \eqref{eq:problem_inf}, it is useful to briefly summarize the Gaussian case,
\begin{equation}
  P = P_0 = \mathcal{N}(\mu_0, \Sigma_0),
\end{equation}
which corresponds to $\varepsilon = 0$. In this case, the MMSE estimator is given by
\begin{equation}
  \label{eq:cond_expectation}
  f_0(Y) \coloneqq E_{P_0}[ X | Y ] = \mu_{X_0} + B_0 C_0^{-1} (Y-\mu_{Y_0}) 
\end{equation}
and the MMSE calculates to
\begin{equation}
   \mmse_{X|Y}(P_0) = \tr\bigl( \Xi_0 \bigr) = \sum_{k=1}^K \xi_{0,k},
\end{equation}
where
\begin{equation}
  \label{eq:schur_complement}
  \Xi_0 \coloneqq \Sigma_0 / C_0 = A_0 - B_0 C_0^{-1} B_0^\text{T}
\end{equation}
denotes the Schur complement of $\Sigma_0$ in $C_0$ and $\xi_{0,1} \geq \xi_{0,2} \geq \ldots \geq \xi_{0,K}$ denote the ordered eigenvalues of $\Xi_0$.

It is now possible to state the main results of this paper.

\begin{theorem}
  For all distributions $P \in \mathcal{P}_\varepsilon(P_0)$, with $P_0$ of the form \eqref{eq:P0_Gaussian}, it holds that 
  \begin{equation}
    \label{eq:mmse_bounds}
    \sum_{k=1}^K \frac{\xi_{0,k}}{1 + \gamma_+ \xi_{0,k}} \leq \mmse_{X|Y}(P) \leq \sum_{k=1}^K \frac{\xi_{0,k}}{1 + \gamma_- \xi_{0,k}},
  \end{equation}
  where $\gamma_+$ and $\gamma_-$ are the unique positive and negative solutions of
  \begin{equation}
    \label{eq:gamma_root}
    \sum_{k=1}^K \phi(\gamma \, \xi_{0,k}) = 2 \varepsilon, \quad \gamma > -\frac{1}{\xi_{0, 1}},
  \end{equation}
  and 
  \begin{equation}
    \phi(t) \coloneqq \log(1+t) - \frac{t}{1+t}, \quad t > -1.
  \end{equation}
  The bounds are attained by the estimator $f_0$ in \eqref{eq:cond_expectation} and the Gaussian distributions $\mathcal{N}(\mu_0, \Sigma_{\gamma_+})$ and $\mathcal{N}(\mu_0, \Sigma_{\gamma_-})$, where
  \begin{equation}
    \Sigma_\gamma = \Sigma_0 - \gamma \begin{bmatrix} \Xi_0 (I_K + \gamma \Xi_0)^{-1} \Xi_0 & 0 \\ 0 & 0 \end{bmatrix},
  \end{equation}
  with $I_K$ being the identity matrix of size $K \times K$.
  \label{th:mmse_bounds}
\end{theorem}

The theorem is proved in the next section. 

\section{Proof of the Main Result}
\label{sec:proof}

The proof of the bounds given in the previous section is based on the Lagrange function
\begin{equation}
    \label{eq:lagrange_function}
    L_\lambda(f, P) \coloneqq \\ \mse_{X | Y}(f, P) + \lambda \DKL{P}{P_0},
\end{equation}
with $\lambda \in \mathbb{R}$. Some useful properties of $L_\lambda$ are stated in the following two Lemmas.

\begin{lemma}
    \label{lm:lagrange_convexity}
    For $P_0 = \mathcal{N}(\mu_0, \Sigma_0)$ and $\varepsilon > 0$
    \begin{enumerate}
        \item $L_\lambda(\bullet, P)$ is strictly convex for all $P \in \mathcal{P}_\varepsilon(P_0)$.
        \item For $\lambda > 0$, $L_\lambda(f, \bullet)$ is strictly convex for all $f \in \mathcal{F}$.
        \item For $\lambda < 0$, $L_\lambda(f, \bullet)$ is strictly concave for all $f \in \mathcal{F}$. 
    \end{enumerate}
\end{lemma}

\begin{lemma}
    \label{lm:px_opt_given_f}
    Let $f \in \mathcal{F}$ be given, let $\eta$ denote the standard Lebesgue measure on $(\mathbb{R}^K, \mathcal{B}^K)$, and let
    \begin{equation}
        \label{eq:def_hf}
        h_f(x,y) \coloneqq \left\lVert x - f(y) \right\rVert^2.
    \end{equation}
    If some $c_f > 0$ exists such that
    \begin{equation}
        \label{eq:pf}
        p_f(x,y) = c_f \, p_0(x,y) \, e^{-\frac{1}{\lambda} h_f(x,y)}
    \end{equation}
    is a valid density w.r.t.~$\eta$, then the corresponding distribution $P_f$ solves
    \begin{equation}
        \inf_{P} \; L_\lambda(f, P), \quad \lambda > 0,
    \end{equation}
    and
    \begin{equation}
        \sup_{P} \; L_\lambda(f, P), \quad \lambda < 0.
    \end{equation}
\end{lemma}

The proofs of both Lemmas follow in close analogy to the proofs of Lemma~1 and Lemma~2 in \cite{DytsoFauss2019} and are hence omitted for brevity. A proof of Lamma~\ref{lm:px_opt_given_f} can also be found in Appendix~A of \cite{Anantharam2018}.

\subsection{Proof of the Lower Bound}
\label{ssec:proof_lower_bound}
First, consider the auxiliary problem
\begin{equation}
    \label{eq:lower_bound_lagrange_problem}
    \inf_f \; \inf_{P} \; L_\lambda(f, P),
\end{equation}
with
\begin{equation}
  \label{eq:lambda}    
  \lambda = \frac{2}{\gamma}, \quad \gamma > 0.
\end{equation}
The inner minimization in \eqref{eq:lower_bound_lagrange_problem} can be solved via Lemma~\ref{lm:px_opt_given_f}:
\begin{align}
   \inf_{P} \; L_\lambda(f, P) &= E_{P_f}\!\left[ h_f(X,Y) + \frac{2}{\gamma} \log \frac{p_f(X,Y)}{p_0(X,Y)} \right] \label{eq:min_L} \\
    &= E_{P_f}\!\left[ h_f(X,Y) + \frac{2}{\gamma} \log c_f - h_f(X,Y) \right] \notag \\
    &= -\frac{2}{\gamma} \log E_{P_0}\Bigl[ e^{-\frac{\gamma}{2} h_f(X,Y)} \Bigr], \label{eq:lower_bound_f}
\end{align}
where the last equality follows from $c_f$ having to be chosen such that $p_f$ is a valid density function. Hence, the optimal estimator in \eqref{eq:lower_bound_lagrange_problem} can be characterized by the problem
\begin{equation}
    \label{eq:lower_bound_f_equiv}
    \sup_f \; E_{P_0}\Bigl[ e^{-\frac{\gamma}{2} h_f(X,Y)} \Bigr].
\end{equation}
The exponential function under the expectation operator is jointly log-concave in $f$ and $x$. Hence, its expected value is log-concave in $f$ so that every stationary point is a global minimum. The G\^ateaux derivative of the objective function in \eqref{eq:lower_bound_f_equiv} in the direction of an estimator $g \in \mathcal{F}$ is given by
\begin{multline}
   -\gamma \, E_{P_0}\Bigl[ \langle X - f(Y) \,,\, g(Y) \rangle e^{-\frac{\gamma}{2} h_f(X,Y)} \Bigr] \\ = -\frac{\gamma}{c_f} E_{P_f} \Bigl[ \langle X - f(Y) \,,\, g(Y) \rangle \Bigr], 
\end{multline}
 where $\langle \bullet \,,\, \bullet \rangle$ denotes the inner product. This yields the necessary and sufficient optimality condition
\begin{equation}
  \label{eq:lower_bound_orthogonality}
  E_{P_f} \Bigl[ \langle X - f(Y) \,,\, g(Y) \rangle \Bigr] \geq 0 
\end{equation}
for all $g \in \mathcal{F}$. Note that this is a classic orthogonality condition, which implies that the optimal estimator is the MMSE estimator under $P_f$, that is,
\begin{equation}
  \label{eq:f_opt_implicit}
  f(Y) = E_{P_f}\bigl[ X | Y \bigr].
\end{equation}

Next, it is shown that this condition is satisfied by the estimator $f_0$ in \eqref{eq:cond_expectation}. In order to see this, note that for $f = f_0$, $P_{f_0}$ in \eqref{eq:pf} is a Gaussian distribution with mean $\mu_{f_0} = \mu_0$ and precision matrix
\begin{equation}
    \Sigma_{f_0}^{-1} = \Sigma_0^{-1} - \gamma U_0^\text{T} U_0
\end{equation}
where
\begin{equation}
    U_0 = \begin{bmatrix} I_K & -B_0 C_0^{-1} \end{bmatrix}. 
\end{equation}
From Woodbury's matrix identity it follows that the corresponding covariance matrix is of the form
\begin{equation}
    \Sigma_{f_0} = \Sigma_0 - \bigl(U_0 \Sigma_0\bigr)^\text{T} \bigl( \tfrac{1}{\gamma}I + U_0 \Sigma_0 U_0^\text{T} \bigr)^{-1} U_0 \Sigma_0.
\end{equation}
Using \eqref{eq:Sigma0_blocks}, 
\begin{align}
    U_0 \Sigma_0 &= \begin{bmatrix} I_K & -B_0 C_0^{-1} \end{bmatrix} \begin{bmatrix} A & B \\ B^\text{T} & C \end{bmatrix} \\
    &= \begin{bmatrix} A_0 - B_0 C_0^{-1} B_0^\text{T}   & B_0 - B_0 C_0^{-1} C_0 \end{bmatrix} \\
    &= \begin{bmatrix} \Xi_0  & 0 \end{bmatrix},
\end{align}
so that $\Sigma_{f_0}$ calculates to
\begin{align}
    \label{eq:lower_bound_sigma_f0}
    \Sigma_{f_0} &= \Sigma_0 - \begin{bmatrix} \Xi_0 (\gamma^{-1} I_K + \Xi_0)^{-1} \Xi_0 & 0 \\ 0 & 0 \end{bmatrix} \\
    &= \Sigma_0 - \gamma \begin{bmatrix} \Xi_0 (I_K + \gamma \Xi_0)^{-1} \Xi_0 & 0 \\ 0 & 0 \end{bmatrix}.
\end{align}
Since the Gaussian MMSE estimator only depends on the mean vector and the $M$ right most columns of the covariance matrix, compare \eqref{eq:cond_expectation}, it immediately follows that
\begin{align}
  E_{P_{f_0}}\bigl[ X | Y \bigr] &= E_{\mathcal{N}(\mu_{f_0}, \Sigma_{f_0})}\bigl[ X | Y \bigr]  \\
  &= E_{\mathcal{N}(\mu_0, \Sigma_0)}\bigl[ X | Y \bigr] \\
  &= f_0(Y),
\end{align}
which is the optimality condition in \eqref{eq:f_opt_implicit}. 

Using this result, it holds that
\begin{align}
    \sup_f \; E_{P_0} &\Bigl[ e^{-\frac{\gamma}{2} \lVert X-f(Y) \rVert_2^2} \Bigr] \\
    &= E_{P_0}\Bigl[ e^{-\frac{\gamma}{2} \lVert (X-\mu_X) - B C^{-1}(Y-\mu_Y) \rVert_2^2} \Bigr] \\
    &= E_{\mathcal{N}(0,I_K)} \Bigl[ e^{ -\frac{\gamma}{2}  \lVert \Xi_0^{1/2} Z \rVert^2  } \Bigr] \\
    &= E_{\mathcal{N}(0,I_K)} \Bigl[e^{ -\frac{\gamma}{2}  \sum_{k=1}^K \xi_{0,k} Z_k^2  }  \Bigr] \label{eq:lower_bound_mgf}
\end{align}
where $Z = [Z_1, \ldots, Z_K]$ is a vector of standard normally distributed random variables. The expression in \eqref{eq:lower_bound_mgf} is the product of $K$ moment generating functions of $\chi^2$ distributed random variables evaluated at $-\frac{\gamma}{2}  \xi_{0,k}$, hence, it evaluates to
\begin{align}
    \prod_{k=1}^K E_{\mathcal{N}(0,1)} \Bigl[e^{ -\frac{\gamma}{2}  \xi_{0,k} Z_k^2  }  \Bigr] &=  \prod_{k=1}^K( 1 + \gamma \xi_{0,k} )^{-\frac{1}{2}}.
\end{align} 
Inserting this result back into \eqref{eq:lower_bound_f} yields
\begin{align}
    \inf_{P_X} \; L_\lambda(f, P_X) &= \frac{1}{\gamma} \sum_{k=1}^K \log( 1 + \gamma \xi_{0,k} )
\end{align}
for all $\gamma > 0$. 

In order to establish the connection to the original problem \eqref{eq:problem_inf}, let $(f^\dagger, P^\dagger)$ denote the solution of the latter. For all $\gamma > 0$ it holds that
\begin{align}
   &\!\!\mmse_{X|Y}(P^\dagger) \notag \\
   &\geq \mse_{X|Y}(f^\dagger,P^\dagger) + \frac{2}{\gamma} \left( \DKL{P^\dagger}{P_0} - \varepsilon \right) \\
   &\geq \inf_{P, f} \left( \mse_{X|Y}(f,P) + \frac{2}{\gamma} \DKL{P}{P_0} \right) - \frac{2}{\gamma} \varepsilon \\
   &\geq \frac{1}{\gamma} \left( \sum_{k=1}^K \log( 1 + \gamma \xi_{0,k} ) - 2 \varepsilon \right) \eqqcolon \rho(\gamma).
\end{align}
Moreover, by strong Lagrange duality \cite[Ch.~5]{Boyd2004},
\begin{equation}
  \mmse_{X|Y}(P^\dagger) = \sup_{\gamma \geq 0} \; \rho(\gamma).
\end{equation}
In order to maximize $\rho(\gamma)$, note that its derivative is given by
\begin{align}
    \rho'(\gamma) &= \frac{1}{\gamma}\left( \sum_{k=1}^K \frac{\xi_{0,k}}{1 + \gamma \xi_{0,k}} - \rho(\gamma) \right) \\
    &\eqqcolon \frac{1}{\gamma}\left( \tilde{\rho}(\gamma) - \rho(\gamma) \right),
\end{align}
where $\tilde{\rho}(\gamma)$ is defined implicitly. Since $\rho$ is concave by construction, every stationary point is a global maximum, which yields the optimality condition
\begin{align}
  \rho(\gamma) -  \tilde{\rho}(\gamma) &= 0 \\
  \sum_{k=1}^K \left( \log( 1 + \gamma \xi_{0,k} ) - \frac{\gamma\xi_{0,k}}{1 + \gamma \xi_{0,k}} \right) &= 2 \varepsilon \\
  \sum_{k=1}^K \phi(\gamma \, \xi_{0,k}) &= 2 \varepsilon.
  \label{eq:opt_condition}
\end{align}
Since $\phi \colon [0, \infty) \to [0, \infty)$ is continuous and increasing, the left-hand side of \eqref{eq:opt_condition} is continuous and increasing in $\gamma$, so that $\gamma_+$ is unique. Finally, by definition of $\gamma_+$,
\begin{equation}
  \rho(\gamma_+) = \tilde{\rho}(\gamma_+) = \sum_{k=1}^K \frac{\xi_{0,k}}{1 + \gamma_+ \xi_{0,k}}.
\end{equation}
This completes the proof.

\subsection{Proof of the Upper Bound}
\label{ssec:proof_upper_bound}
The proof of the upper bound follows by considering the auxiliary problem 
\begin{equation}
    \label{eq:upper_bound_lagrange_problem}
    \inf_f \; \sup_{P} \; L_\lambda(f, P),
\end{equation}
with $\lambda$ as in \eqref{eq:lambda} and $\gamma < 0$. Following the steps of the proof of the lower bound results in the optimality condition \eqref{eq:opt_condition} with $\gamma < 0$. Since $\phi \colon (-1, 0] \to [0, \infty)$ is continuous and decreasing, $\gamma_-$ is again unique and $\rho(\gamma_-)$ evaluates to the right-hand side of \eqref{eq:mmse_bounds}. This completes the proof.

\section{Discussion}
\label{sec:discussion}

In analogy to the MMSE bounds in \cite{DytsoFauss2019}, the upper bound in Theorem~\ref{th:mmse_bounds} provides a robustness result, while the lower bound provides a fundamental limit on the estimation accuracy. In particular, the lower bound is a useful alternative to well-known inequalities in estimation and information theory, such as the Cram\'{e}r--Rao lower bound, Stam's inequality, or the entropy power inequality. Naturally, the quality of the proposed bound depends on how well the distribution $P$ can be approximated by a Gaussian distribution. However, since the requirement of having a finite KL divergence to a Gaussian reference distribution is relatively mild, it is defined for a larger class of input distributions than the Cram\'{e}r--Rao bound.

\subsection{Minimax Robust MMSE Estimation}
The upper bound provides a minimax result in the sense that for all $P \in \mathcal{P}_\varepsilon(P_0)$, the MSE of the estimator $f_0$ is guaranteed to be bounded and that $f_0$ minimzes this bound. That is, any estimator $f \neq f_0$ can only deteriorate the worst case performance over the set $\mathcal{P}_\varepsilon(P_0)$. This leads to the following, somewhat surprising, corollary.

\begin{corollary}[Minimax robustness of linear estimators]
  Every linear estimator
  \begin{equation}
    f_\text{lin}(Y) = a + H (Y - b),
  \end{equation}
  with $a \in \mathbb{R}^K$, $b \in \mathbb{R}^M$, $H \in \mathbb{R}^{K \times M}$ is minimax optimal with respect to the MMSE under distributional uncertainty of the KL divergence ball type. More formally, $f_\text{lin}$ satisfies
  \begin{equation}
    \sup_{P \in \mathcal{P}_\varepsilon(H)} \; \mse_{X|Y}(f_\text{lin}, P) \leq \sup_{P \in \mathcal{P}_\varepsilon(H)} \; \mse_{X|Y}(f, P)
  \end{equation}
  for all $f \in \mathcal{F}$, where
  \begin{equation}
    \mathcal{P}_\varepsilon(H) \coloneqq \bigcup_{\Sigma_0 \in \mathcal{S}_H} \mathcal{P}_\varepsilon\bigl(\mathcal{N}(\mu_0, \Sigma_0)\bigr),
  \end{equation}
  with $\mu_0$ as in \eqref{eq:mu0} and
  \begin{multline}
    \mathcal{S}_H \coloneqq \biggl\{ \Sigma \in \mathbb{S}_+^{K+M} : \Sigma = \begin{bmatrix} A & HC \\ (HC)^\text{T} & C \end{bmatrix}, \\ A \in \mathbb{S}^{K}, C \in \mathbb{S}_+^{M} \biggr\}  .
  \end{multline}
  \label{cor:lin_est_robustness}
\end{corollary}
Corollary~\ref{cor:lin_est_robustness} follows immediately from Theorem~\ref{th:mmse_bounds} by observing that all $\Sigma_0 \in \mathcal{S}_H$ correspond to the same MMSE estimator $f_0 = f_{\text{lin}}$, which in turn solves the problem in \eqref{eq:problem_sup}.

The result in Corollary~\ref{cor:lin_est_robustness} is counter intuitive at first glance since linear estimators are well-known \emph{not} to be robust against distributional uncertainty; see \cite{Zoubir2012, Zoubir2018} and references therein. In particular, every linear estimator admits an unbounded influence function \cite{Hampel1986, Asato1992}, meaning that the error caused by a single outlier can be arbitrarily large. The minimax result in Corollary~\ref{cor:lin_est_robustness} does not contradict these findings. First, it states a bound on the \emph{expected} square error of the estimator, so that individual estimates might still be highly inaccurate in some unlikely cases. Second, the KL divergence ball uncertainty model implicitly limits the probability of extreme outliers since distributions with heavy tails also admit large KL divergences with respect to a Gaussian distribution. In this sense, Corollary~\ref{cor:lin_est_robustness} states that linear estimators are insensitive against distributions only being \emph{approximately} Gaussian, but not against distributions admitting drastically different tail behavior. 

\subsection{Additive Noise Channels}
The bounds presented in \cite{DytsoFauss2019} hold for additive noise channels in which noise and input are independent and at least one of them is Gaussian distributed. With the bounds in Theorem~\ref{th:mmse_bounds} at hand, these assumptions can be relaxed. Moreover, if the additive noise channel is approximated by a additive Gaussian noise (AGN) channel, the KL divergence of the joint input-output distributions simplifies to the sum of the KL divergences of the input and the noise distributions. 

In general, the KL divergence between two distributions $P_{XY} = P_X P_{Y|X}$ and $Q_{XY} = Q_X Q_{Y|X}$ can be decomposed into
\begin{multline}
  \DKL{P_{XY}}{Q_{XY}} \\ = \DKL{P_{X}}{Q_{X}} + E_{P_X}\bigl[ \DKL{P_{Y|X}}{Q_{Y|X}} \bigr].
  \label{eq:KL_decomposition}
\end{multline}
Now, consider an additive channel
\begin{equation}
    Y = X + N,
    \label{eq:AGN}
\end{equation}
where $X \sim P_X$ and $N \sim P_N$ are independent. In this case it holds that
\begin{align}
  E_{P_X}\bigl[ \DKL{P_{Y|X}}{Q_{Y|X}} \bigr] &= E_{P_X}\bigl[ \DKL{P_{Y-X}}{Q_{Y-X}} \bigr] \notag \\
  &= E_{P_N}\bigl[ \DKL{P_{N}}{Q_{N}} \bigr] \\
  &= \DKL{P_{N}}{Q_{N}},
\end{align}
so that
\begin{align}
    \DKL{P_{XY}}{Q_{XY}} &= \DKL{P_X}{Q_X} + \DKL{P_N}{Q_N}.
    \label{eq:KL_additive}
\end{align}
Moreover, if the reference distribution $P_0$ is chosen such it corresponds to an additive Gaussian noise channel 
\begin{equation}
  Y_0 = X_0 + N_0,
\end{equation}
where $X_0 \sim \mathcal{N}(\mu_{X_0}, \Sigma_{X_0})$ and $N_0 \sim \mathcal{N}(\mu_{N_0}, \Sigma_{N_0})$, then the MMSE matrix $\Xi_0$ in \eqref{eq:schur_complement} simplifies to
\begin{equation}
    \Xi_0 = \Sigma_{X_0} (\Sigma_{X_0} + \Sigma_{N_0})^{-1} \Sigma_{N_0}.
    \label{eq:MMSE_additive}
\end{equation}

Using these results, MMSE bounds for additive noise channels can be obtained by adding the non-Gaussianity parameters of the input and noise distributions instead of considering their joint non-Gaussianity. This is a natural extension of the bounds in \cite{DytsoFauss2019} and will be illustrated with an example in the next section. However, this simplicity comes at the cost of less tight bounds, since only allowing reference distributions that correspond to AGN channels reduces the degrees of freedom.

\subsection{A Special Case with Explicit Bounds}
\begin{figure}[tb]
  \centering
  \includegraphics{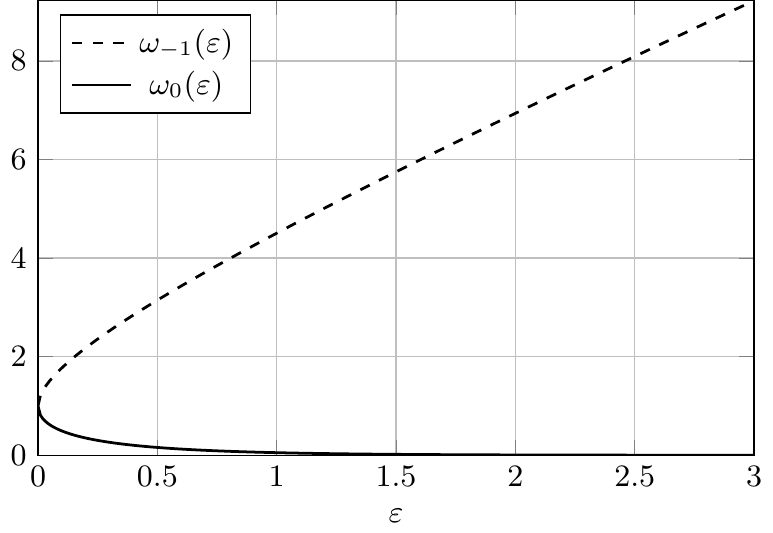}
  \caption{Graphs of $\omega_{-1}$ and $\omega_0$ defined in \eqref{eq:w}.}
  \label{fig:w}
\end{figure}

For the special case that the covariance matrix of the reference distribution, $\Sigma_0$, is chosen such that its Schur complement admits a flat spectrum, that is, if $\xi_{0,1} = \ldots = \xi_{0,K} = \xi_0$, the solution of \eqref{eq:gamma_root} can be expressed explicitly, namely
\begin{align}
  \gamma_+ \xi_{0} = \frac{1-\omega_{0}(\varepsilon/K)}{\omega_{0}(\varepsilon/K)} \quad \text{and} \quad \gamma_- \xi_{0} = \frac{1-\omega_{-1}(\varepsilon/K)}{\omega_{-1}(\varepsilon/K)},
\end{align}
where
\begin{equation}
  \omega_i(t) = -W_i\bigl(-e^{-(2t+1)}\bigr)
  \label{eq:w}
\end{equation}
and $W_i$ denotes the $i$th branch of the Lambert W function \cite{Corless1996}. Inserting \eqref{eq:w} back into \eqref{eq:mmse_bounds} yields bounds of the simple form
\begin{equation}
    \label{eq:mmse_bounds_scalar}
    \omega_0(\varepsilon/K) \leq \frac{\mmse_{X|Y}(P)}{\mmse_{X|Y}(P_0)} \leq \omega_{-1}(\varepsilon/K),
\end{equation}
where $\mmse_{X|Y}(P_0) = K \xi_0$. For illustration purposes, the functions $\omega_{-1}$ and $\omega_0$ are plotted in Fig.~\ref{fig:w}. As can be seen, $\omega_0$ quickly approaches zero, meaning that the lower bound is most useful for distributions that are sufficiently close to a Gaussian distribution. On the other hand, $\omega_{-1}$ grows approximately linearly in $\varepsilon$, which means that the upper bound can potentially be useful for a larger class of distributions. This behavior is also interesting in light of Corolarry~\ref{cor:lin_est_robustness} since it indicates the rate at which the worst case MSE of a linear estimator increases with the non-Gaussianity measure $\varepsilon$ is approximately constant---this can also be observed in a more realistic example presented in the next section. 

Moreover, \eqref{eq:mmse_bounds_scalar} indicates that the proposed bounds are asymptotically exact for $K \to \infty$, whenever the KL divergence between $P$ and $P_0$ grows sublinearly in $K$, that is $\DKL{P}{P_0} \in o(K)$. This is in line with the results in \cite{DytsoFauss2019}, where this behavior was demonstrated for uniform input distributions on balls in $\mathbb{R}^K$.

\subsection{Bounds on $\gamma_+$ and $\gamma_-$}
In order to solve \eqref{eq:gamma_root} for $\gamma$, it is useful to be able to bound $\gamma_+$ and $\gamma_-$ from above and below, so that the problem can be reduced to finding the root of a monotonic function on a finite interval. The following corollary provides such bounds

\begin{corollary}
  For $\gamma_+$ and $\gamma_-$ as in Theorem~\ref{th:mmse_bounds} it holds that
  \begin{align}
      \frac{1-\omega_{0}(\varepsilon/K)}{\omega_{0}(\varepsilon/K)} &\leq \gamma_+ \xi_{0,1} \leq \frac{1-\omega_{0}(\varepsilon)}{\omega_{0}(\varepsilon)}, \\
      \frac{1-\omega_{-1}(\varepsilon)}{\omega_{-1}(\varepsilon)} &\leq \gamma_- \xi_{0,1}  \leq \frac{1-\omega_{-1}(\varepsilon/K)}{\omega_{-1}(\varepsilon/K)},
  \end{align}
  with $\omega_i$ defined in \eqref{eq:w}.
\end{corollary}

The corollary follows in a straightforward manner from the monotonicity of $\phi$ and the bounds
\begin{equation}
    \phi(\gamma \, \xi_{0,1}) \leq \sum_{k=1}^K \phi(\gamma \, \xi_{0,k}) \leq K \phi(\gamma \, \xi_{0,1}).
\end{equation}

\subsection{Connection to the AWGN Channel}
A possibly helpful interpretation of the bounds in Theorem~\ref{th:mmse_bounds} is via the AWGN channel:
\begin{equation}
  \label{eq:channel_snr}
  U = \sqrt{\gamma} X + N,
\end{equation}
where $N \sim \mathcal{N}(0,I_K)$, $X \sim \mathcal{N}(\mu_X, \Xi_0)$, and $\gamma$ denotes the signal-to-noise ratio (SNR). The MMSE of the channel in \eqref{eq:channel_snr} is given by
\begin{equation}
  \mmse_{X|U}(\gamma) = \sum_{k=1}^K \frac{\xi_{0,k}}{1 + \gamma \xi_{0,k}},
\end{equation}
so that the bounds in \eqref{eq:mmse_bounds} can be written as
\begin{align}
  \mmse_{X|U}(\gamma_+) \leq \mmse_{X|Y}(P) \leq \mmse_{X|U}(\gamma_-).
\end{align}
Interestingly, the upper bound in \eqref{eq:mmse_bounds} corresponds to a negative SNR value, which does not admit an obvious physical interpretation. 

\section{Examples}
\label{sec:examples}

In this section, it is outlined how the presented bounds can be applied in different contexts. Python code for the bounds and all examples can be found in a public Git repository \cite{GitHubRepo}. 

The main mechanism behind the application of the bounds is the idea of  the best Gaussian approximation.  Specifically, for a given joint distribution $P_{XY}$ on $(X,Y)$,  we lower bound the MMSE by 
\begin{align}
&\!\!\mmse_{X|Y}(P_{XY})\notag\\
&\ge \sup_{P_0} \inf_{P} \Big\{ \mmse_{X|Y}(P)  : P \in \mathcal{P}_{D_\text{KL}(P_{XY}\Vert P_0)}(P_0)  \Big\},
\end{align}
and upper bound the MMSE by 
\begin{align}
&\!\!\mmse_{X|Y}(P_{XY}) \notag\\
&\le \inf_{P_0} \sup_{P} \Big\{ \mmse_{X|Y}(P) : P \in \mathcal{P}_{D_\text{KL}(P_{XY}\Vert P_0)}(P_0)  \Big\},
\end{align}
where in both cases $P_0$ is restricted to be Gaussian.

\subsection{Minimax Robust MMSE Estimation}
The upper MMSE bound in Theorem~\ref{th:mmse_bounds} provides a robustness result for a Gaussian nominal model under missmatch of the KL divergence ball type detailed in Corollary~\ref{cor:lin_est_robustness}. In order to evaluate the corresponding robust estimator, typically one would compare the latter to a suitable non-robust estimator in terms of worst case and nominal performance. Here, however, the minimax estimator and the nominal estimator in fact coincide, so that this comparison becomes unnecessary. Put another way, the minimax property under KL divergence uncertainty comes for free when using the standard linear MMSE estimator.

\begin{figure}[tb]
  \centering
  \includegraphics{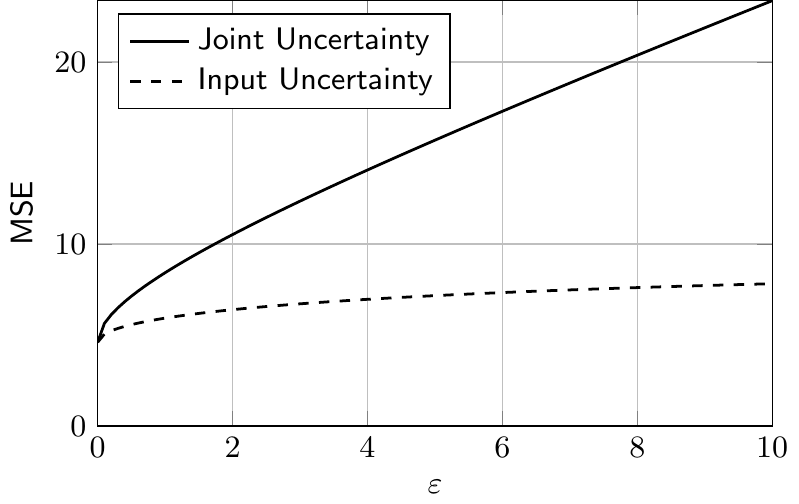}
  \caption{Minimax MMSE for a correlated signal in white Gaussian noise at \SI{0}{\decibel} SNR when the joint distribution or only the input distribution are contained in a KL divergence ball with radius $\varepsilon$.}
  \label{fig:mse_vs_eps}
\end{figure}

Nevertheless, it is instructive to compare the minimax result in Corollary~\ref{cor:lin_est_robustness} to the minimax result in \cite{DytsoFauss2019}, where only the input distribution was subject to uncertainty. To this end, consider the example in \cite{DytsoFauss2019}, Section~5.A, where a Gaussian signal with $K = 10$ and input covariance matrix
\begin{equation}
  [\Sigma_X]_{ij} = e^{-0.9 \lvert i - j \rvert }, \quad i,j = 1, \ldots, 10.
  \label{eq:covariance_example}
\end{equation} 
is estimated in white Gaussian noise with variance $\sigma_N^2 = 1/\gamma$ so that the SNR is given by $\tr(\Sigma_X)/\tr(\Sigma_N)$. The two minimax MSEs corresponding to the two uncertainty models are plotted in Fig.~\ref{fig:mse_vs_eps}. As can be seen, the difference in the best attainable worst case performance is stark. Note that this effect is not due to one model allowing for more uncertainty. Constraining the joint distribution instead of the input distribution to be in a KL divergence ball of radius $\varepsilon$ does not introduce more uncertainty \emph{per se}, but, according to \eqref{eq:KL_additive}, allows for it to be distributed between the input distribution and the channel. Naturally, the channel is much more sensitive to deviations from the nominal case. In other words, tempering with the channel offers a hypothetical adversary much more leverage than tempering with the input distribution only.  

\begin{figure}[tb]
  \centering
  \includegraphics{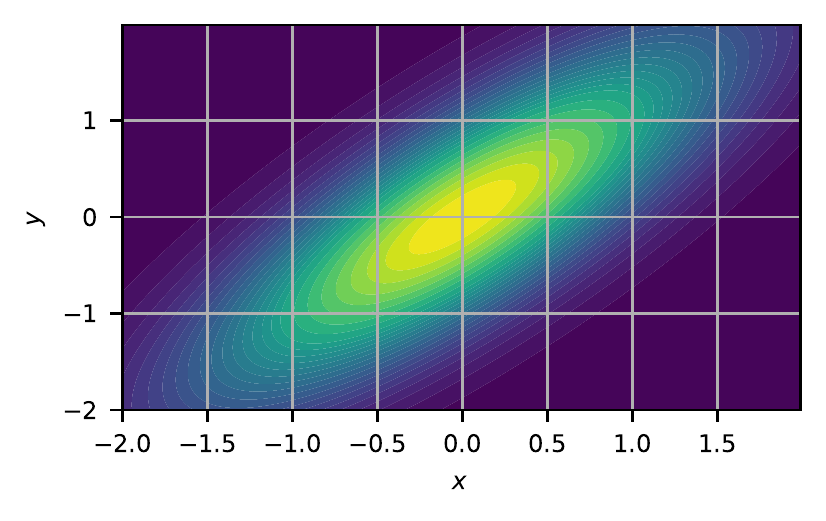}
  \includegraphics{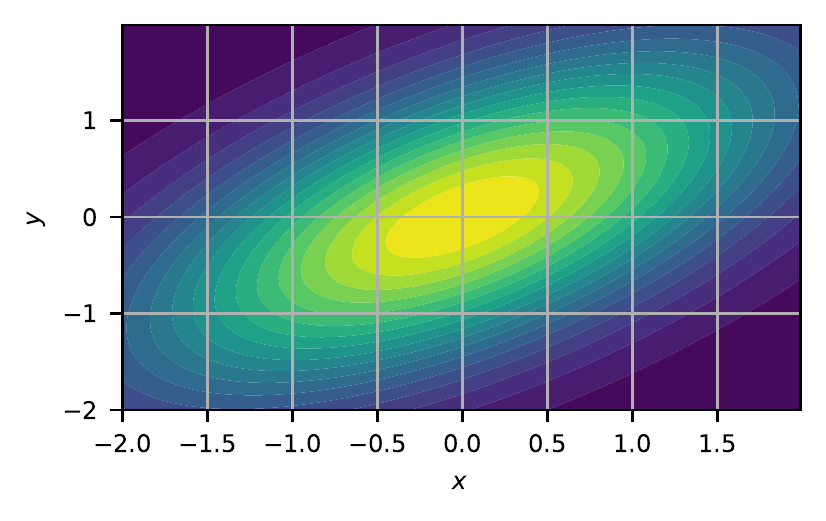}
  \includegraphics{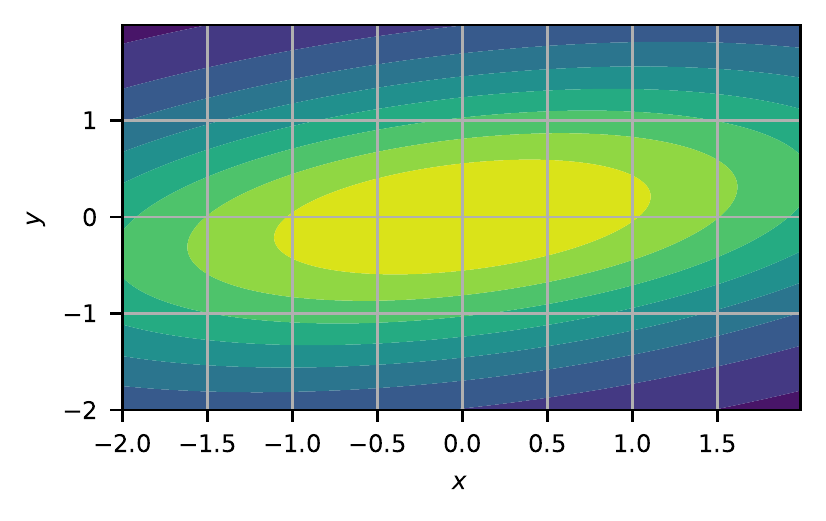}
  \caption{Illustration of least favorable joint distributions for an additive Gaussian noise channel with $K = M = 1$ at \SI{3}{\decibel} SNR under varying degrees of uncertainty: $\varepsilon = 0$ (top), $\varepsilon = 0.5$ (middle), $\varepsilon = 5$ (bottom).}
  \label{fig:lfd_vs_eps}
\end{figure}

This effect is illustrated in Fig~\ref{fig:lfd_vs_eps}, where the least favorable distributions for different KL ball radii are shown. For the sake of a graphical representation, here the scalar case is considered, $K=1$, with $\sigma_X^2 = 1$ and an SNR of \SI{3}{\decibel}. From top to bottom the KL divergence ball radius was chosen to be $\varepsilon = 0$ (nominal model), $\varepsilon = 0.5$ (slight mismatch), and $\varepsilon = 5$ (severe mismatch). It can clearly be seen how not only the signal power, but, more importantly, also the correlation between input and output changes when the uncertainty is increased. From a robustness perspective, this possibility of increasing the input power while at the same time reducing the correlation with the output makes uncertainty in the joint distribution much more critical.

\subsection{Estimating a Generalized-Gaussian Signal in Generalized-Gaussian Noise}
In \cite{DytsoFauss2019}, MMSE bounds for estimating a generalized Gaussian (GG) signal in additive Gaussian noise were presented. Here, this example is generalized to a GG signal in GG noise. Using the channel model in \eqref{eq:AGN}, let $X \sim \mathcal{G}(a, p)$ and $N \sim \mathcal{G}(b, q)$, where $\mathcal{G}(a, p)$ denotes a generalized Gaussian distribution with density function
\begin{equation}
    g(x | a, p) = \frac{p}{2a\Gamma(1/p)} e^{-\left( \frac{\lvert x \rvert}{a}\right)^p},
\end{equation}
where $\Gamma$ denotes the gamma function \cite{Davis1959}, $a > 0$ is a scale parameter, and $p > 0$ determines the type of decay of the tails \cite{DytsoISIT2017}. In \cite{DytsoFauss2019}, it is shown that the best Gaussian approximation of a zero-mean GG distribution, in terms of the KL divergence, is attained by choosing the variance of the reference distribution as
\begin{equation}
    \sigma_0^2 = a \sqrt{\frac{\Gamma(3/p)}{\Gamma(1/p)}},
\end{equation}
so that
\begin{align}
    d_\mathcal{G}(p) &\coloneqq \min_{\sigma_0^2 \geq 0} \; \DKL{\mathcal{G}(a,p)}{\mathcal{N}(0,\sigma_0^2)} \label{eq:I-Projetion} \\
  &= \log \frac{p}{\sqrt{2}} \sqrt{ \frac{\Gamma(3/p)}{\Gamma(1/p)}}\frac{\Gamma(1/2)}{\Gamma(1/p)}  + \frac{1}{2} - \frac{1}{p}. \label{eq:dG}
\end{align}
See Fig.~4 in \cite{DytsoFauss2019} for a plot of the graph of $d_\mathcal{G}$. From \eqref{eq:KL_additive} it follows that the KL divergence of the true input-output distribution\footnote{Note that unless both the input and the noise are Gaussian distributed, the joint input-output distribution is not a multivariate GG distribution itself.} and its best (additive) Gaussian approximation is given by
\begin{equation}
    d_{XY}(p,q) \coloneqq d_\mathcal{G}(p) + d_\mathcal{G}(q).
    \label{eq:gg_non_gaussianity}
\end{equation}
Combining \eqref{eq:gg_non_gaussianity}, \eqref{eq:MMSE_additive}, and \eqref{eq:mmse_bounds_scalar} yields the bound
\begin{align}
  \mmse_{Y|X}(P_\mathcal{GG}) &\geq \omega_0(d_{X,Y}(p,q)) \mmse_{Y|X}(P_0) \\
  &= \omega_0(d_{X,Y}(p,q)) \frac{\sigma_X^2 \sigma_N^2}{\sigma_X^2 + \sigma_N^2},
  \label{eq::gg_lower_bound}
\end{align}
where $\sigma_X^2$ and $\sigma_N^2$ denote the signal and noise power, respectively. 

\begin{figure}[tb]
    \centering
    \includegraphics{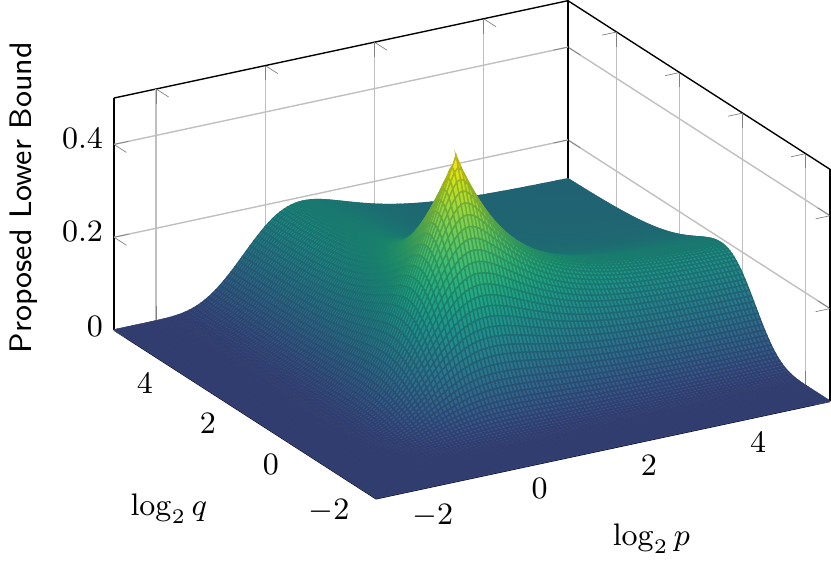}
    \includegraphics{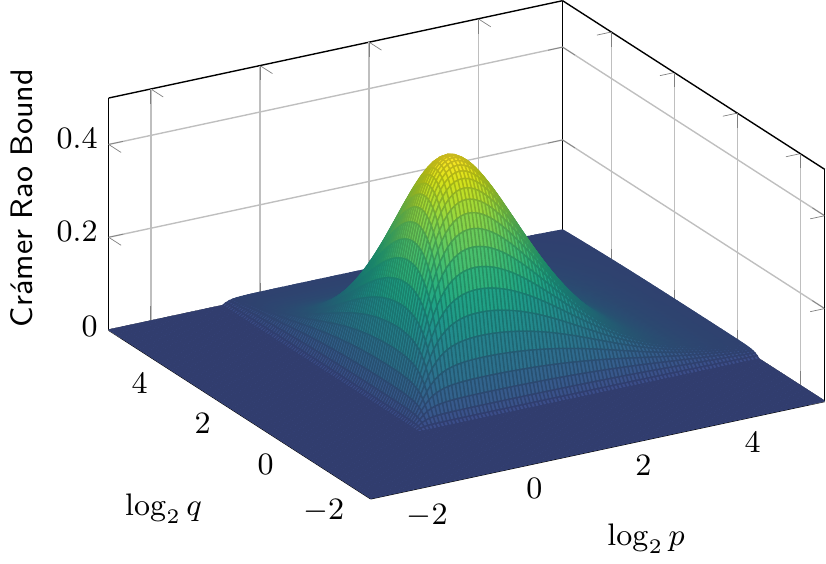}
    \caption{Proposed MMSE lower bound (top) and Cram\'er--Rao bound (bottom) for an additive channel with generalized Gaussian noise and input.}
    \label{fig:gg_bounds}
\end{figure}

Examples of the lower bound in \eqref{eq::gg_lower_bound} are shown in the upper plot of Fig.~\ref{fig:gg_bounds} for $p, q \in [2^{-2}, 2^5]$ and at an SNR of \SI{0}{\decibel} ($\sigma_X^2 = \sigma_N^2 = 1$). For comparison, the Cr\'amer Rao bound (CRB) is depicted in the lower plot. The latter can be shown to be given by
\begin{equation}
  \mmse_{Y|X}(P_\mathcal{GG}) \geq \frac{1}{I\bigl( \mathcal{G}(a,p) \bigr) + I\bigl( \mathcal{G}(b,q) \bigr)},
  \label{eq:Bayesian_CR_GG}
\end{equation}
where
\begin{equation}
  I\bigl( \mathcal{G}(a,p) \bigr) = \left\{ \begin{array}{ll}   \frac{p^2}{a^2} \frac{\Gamma(2-1/p)}{\Gamma(1/p)},  &1/2 <p<\infty \\
  \infty ,    &0 < p\le 1/2 
  \end{array}  \right.   \label{eq:gg_fisher_information}
\end{equation}
denotes the Fisher information of the zero-mean generalized Gaussian distribution \cite[Chapter 3.2.1]{Kassam_2012}.

By inspection, the lower bound proposed here is an improvement over the CRB for a variety of combinations of $p$ and $q$. In particular, the proposed bound is significantly tighter as long as one of the distributions is close to Gaussian ($p,q \approx 2$), while the other distribution is more concentrated ($q,p > 2$). In contrast, it can be seen that the CRB only performs well if both distributions are approximately Gaussian, with a pronounced peak around $\log_2 p = \log_2 q = 1$.

\begin{figure}[tb]
    \centering
    \includegraphics{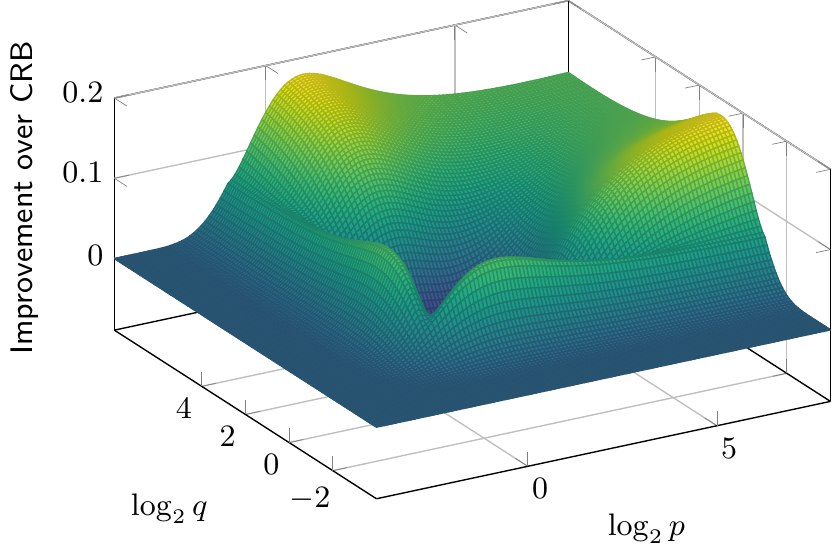}
    \includegraphics{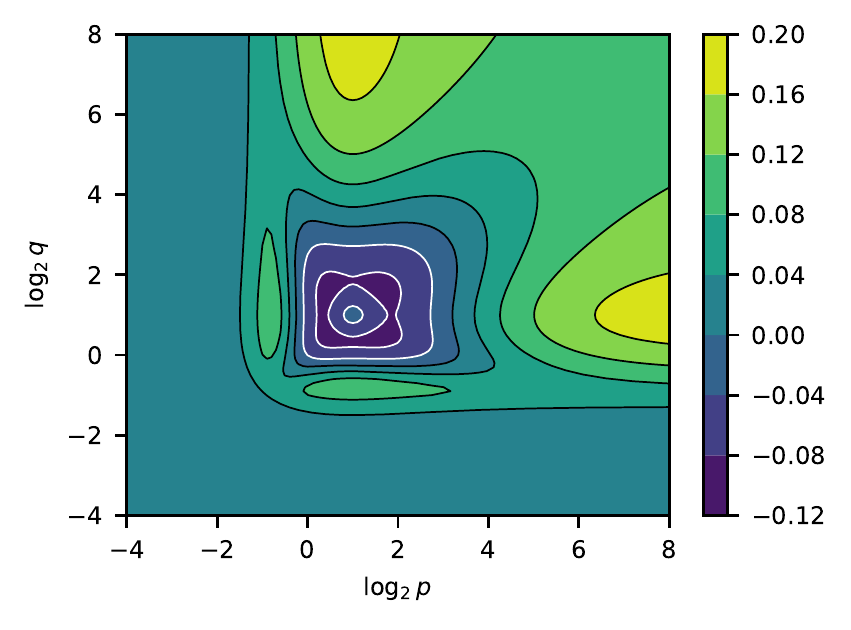}
    \caption{Difference between the proposed MMSE lower bound and the Cram\'er--Rao bound for an additive channel with generalized Gaussian noise and input.}
    \label{fig:delta_bounds}
\end{figure}

This improvement becomes more obvious when considering the difference between the two bounds, which is plotted in Fig.~\ref{fig:delta_bounds}. Again, the proposed bound is notably tighter, with the exception of a region around the Gaussian case. Since this region is difficult to recognize in the surface plot, it is shown separately in the plot below, where it is indicated by the white contour lines.

\subsection{Multiplicative Channel with Uniform Input Distribution}
It is also instructive to extend another example from \cite{DytsoFauss2019} to the non-Gaussian noise case, namely that of uniform input distributions on $K$-dimensional balls, $K$-balls for short. The letter is defined as
\begin{equation}
  \mathcal{B}_K(c, r) = \left\{ x \in \mathbb{R}^K : \sum_{k=1}^K \frac{(x_k-c_k)^2}{r^2} \leq 1 \right\},
\end{equation}
where $r > 0$ denotes the radius of the $K$-ball and $c \in \mathbb{R}^K$ denotes its center. 

Now, consider the multiplicative channel
\begin{equation}
  Y = X \cdot N,
  \label{eq:mulitplicative_channel}
\end{equation}
where $\cdot$ denotes the elementwise product, $N \in \mathbb{R}^K$ is standard normally distributed, $N \sim \mathcal{N}(0, I_K)$, and $X$ is uniformly distributed on $\mathcal{B}_K(c, r)$, here denoted by $X \sim \mathcal{U}_{\mathcal{B}_K}(c, r)$. For simplicity, it is assumed that $\mathcal{B}_K(c, r) \subset \mathbb{R}_+^K$, that is, $c_k > r$ for all $k = 1, \ldots, K$. 

The joint distribution of $X$ and $Y$ in \eqref{eq:mulitplicative_channel} can be approximated by jointly Gaussian random variables $X_0$ and $Y_0$ as follows. First, it is shown in \cite{DytsoFauss2019} that the best Gaussian approximation for $\mathcal{B}_K(c, r) \subset \mathbb{R}_+^K$ is obtained by moment matching, that is,
\begin{equation}
  \mu_{X_0} = c \quad \text{and} \quad \Sigma_{X_0} = I_k \frac{r^2}{K+2}.
  \label{eq:uniform_gaussian_input}
\end{equation}
The KL divergence between $P_X = \mathcal{U}_{\mathcal{B}_K}(c, r)$ and $P_{X_0} = \mathcal{N}(\mu_{X_0, \sigma_{X_0}^2})$ can be shown to be given by
\begin{align}
  \DKL{P_X}{P_{X_0}} &= \frac{K}{2} - \frac{K}{2}\log\frac{K+2}{2} + \log \Gamma\frac{K+2}{2} \\
  &\eqqcolon d_\mathcal{U}(K). \label{eq:kl_uniform}
\end{align}
Since in a jointly Gaussian channel the conditional variance of $Y|X$ is independent of $X$, the conditional distribution $P_{Y|X} = \mathcal{N}(0, I_K X^2)$ can only be approximated by a Gaussian distribution with fixed, diagonal covariance matrix, so that $P_{Y_0|X_0} = P_{Y_0} = \mathcal{N}(0, I_K\sigma_{Y_0}^2)$. The corresponding conditional KL divergence is given by
\begin{equation}
  \DKL{P_{Y|X}}{P_{Y_0}} = \sum_{k=1}^K \DKL{P_{Y_k|X_k}}{P_{Y_{0,k}}},
\end{equation}
where
\begin{equation*}
 \DKL{P_{Y_k|X_k}}{P_{Y_{0,k}}} = \frac{1}{2} \left( \frac{X_k^2}{\sigma_{Y_{0,k}}^2} - 1 -  \log \frac{X_k^2}{\sigma_{Y_{0,k}}^2} \right).
\end{equation*}
In order to evaluate $\DKL{P_{XY}}{P_{X_0Y_0}}$ via \eqref{eq:KL_decomposition}, the expected value of $\DKL{P_{Y|X}}{P_{Y_0}}$ with respect to $P_X$ is required, which is given by
\begin{align}
  &\!E_{P_X} \bigl[ \DKL{P_{Y|X}}{P_{Y_0}} \bigr] \\
  &= \sum_{k=1}^K E_{P_{X_k}} \bigl[ \DKL{P_{Y_k|X_k}}{P_{Y_{0,k}}} \bigr] \\
  &= \frac{1}{2} \sum_{k=1}^K \left( \frac{E[X_k^2]}{\sigma_{Y_{0,k}}^2} - 1 -  E\bigl[ \log X_k^2 \bigr] + \log \sigma_{Y_{0,k}}^2 \right).
\end{align} 
Minimizing with respect to $\sigma_{Y_0|X_0}^2$ yields the best Gaussian approximation $\sigma_{Y_{0,k}}^2 = E[X_k^2]$, so that
\begin{multline}
  E_{P_X} \bigl[ \DKL{P_{Y|X}}{P_{Y_0}} \bigr] \\
  = \frac{1}{2} \sum_{k=1}^K \left( \log E\bigl[ X_k^2 \bigr] - E\bigl[ \log X_k^2 \bigr] \right).
\end{multline}
It is not difficult to show\footnote{For a unit $K$-ball centered at the origin, the probability of the event $\{X_k \leq x\}$, $x \in [0,1]$, corresponds to the ratio of the volume of the spherical cap \cite{Li2011} of height $x$ to the volume of the entire $K$-ball.} that
\begin{equation}
  p_{X_k}(x) =  \frac{1}{\sqrt{\pi} \, r} \frac{\Gamma(\frac{K+2}{2})}{\Gamma(\frac{K+3}{2})} \, \beta_{1, \frac{K+1}{2}}\biggl( \frac{(x - c_k)^2}{r^2}\biggr),
  \label{eq:pdf_xk}
\end{equation}
where $\beta_{a,b}$ denotes the PDF of the beta distribution with parameters $a$ and $b$. From \eqref{eq:pdf_xk} it follows that 
\begin{equation}
  E\bigl[ X_k^2 \bigr] = c_k^2 + \frac{r^2}{K+2}
\end{equation}
and
\begin{equation}
  E\bigl[ \log X_k^2 \bigr] = \log c_k^2 + \frac{2}{\sqrt{\pi}} \frac{\Gamma(\frac{K+2}{2})}{\Gamma(\frac{K+1}{2})} \, H\Bigl(\frac{r}{c_k}, \frac{K+1}{2}\Bigr),
  \label{eq:kl_mult_channel}
\end{equation}
where $H \colon (0,1] \times \mathbb{R}_+ \to \mathbb{R}$ is defined as
\begin{equation}
  H(a,b) \coloneqq \int_{-1}^{1} \log(1+ax) (1-x^2)^{b-1} \, \mathrm{d}x.
\end{equation}
If evaluating the right hand side of \eqref{eq:kl_mult_channel} is too costly, the bound
\begin{align}
  E\bigl[ \log X_k^2 \bigr] &> \frac{1}{2} \left( \log(c_k-r)^2 + \log(c_k+r)^2 \right) \\
  &= \log(c_k^2 - r^2)
  \label{eq:eps_approx}
\end{align}
can be used instead, which is obtained by lower bounding the logarithmic function by an affine function on the interval $[c_k-r, c_k+r]$ and is a good approximation when $c_k \gg r$.

Given this Gaussian approximation and using the fact that $\Sigma_0$ in \eqref{eq:uniform_gaussian_input} admits a flat spectrum, the lower MMSE bound in Theorem~\ref{th:mmse_bounds} evaluates to
\begin{align}
  \mmse_{X|Y}(P_{XY}) &\geq K \omega_{-1}\biggl(\frac{\varepsilon_K(c,r)}{K} \biggr) \sigma_{X_{0,1}}^2 \\
  &= \omega_{-1}\biggl(\frac{\varepsilon_K(c,r)}{K} \biggr) \frac{K}{K+2}r^2, 
  \label{eq:uniform_lower_bound}
\end{align}
where 
\begin{align}
  \varepsilon_K(c,r) &= \DKL{P_{XY}}{P_{X_0 Y_0}} \\
  &= d_\mathcal{U}(K) + \frac{1}{2} \sum_{k=1}^K d_\beta(c_k, r, K),
  \label{eq:eps_uniform}
\end{align}
with $d_\mathcal{U}$ defined in \eqref{eq:kl_uniform} and 
\begin{align}
  d_\beta(c_k, r, K) &= \log E\bigl[ X_k^2 \bigr] - E\bigl[ \log X_k^2 \bigr] \\
  &= c_k^2 + \frac{r^2}{K+2} - \log c_k^2 \notag \\
  & \quad - \frac{2}{\sqrt{\pi}} \frac{\Gamma(\frac{K+2}{2})}{\Gamma(\frac{K+1}{2})} \, H\Bigl(\frac{r}{c_k}, \frac{K+1}{2}\Bigr) \\
  &< c_k^2 + \frac{r^2}{K+2} -  \log(c_k^2 - r^2).
\end{align}
The upper MMSE bound is given by
\begin{equation}
  \mmse_{X|Y}(P_{XY}) \leq \frac{K}{K+2} r^2,
  \label{eq:uniform_upper_bound}
\end{equation}
which can be obtained immediately from $\mmse_X(P_X) = \text{Var}(X) > \mmse_{X|Y}(P_{XY})$ or by minimizing the upper bound proposed here w.r.t.~$\sigma_{X_0}^2$; compare Section 5.B in \cite{DytsoFauss2019}.

\begin{figure}[tb]
    \centering
    \includegraphics{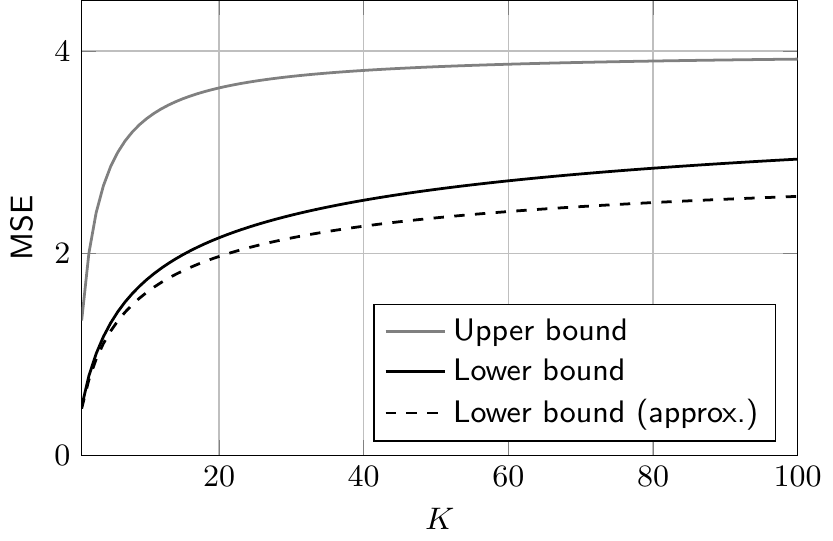}
    \caption{Upper and lower bound on the MMSE of the model in \eqref{eq:mulitplicative_channel}, where $X$ is distributed uniformly on a $K$-ball with center point $c_1 = \ldots = c_K = 10$ and radius $r = 2$. The upper bound is given in \eqref{eq:uniform_upper_bound}, the lower bound in  \eqref{eq:uniform_lower_bound}, and the approximate lower bound is obtained by using the inequality in \eqref{eq:eps_approx} to bound $\varepsilon_K$ in \eqref{eq:eps_uniform}.} 
    \label{fig:uniform_bounds}
\end{figure}

An example of the bounds in \eqref{eq:uniform_lower_bound} and \eqref{eq:uniform_upper_bound} is shown in Fig.~\ref{fig:uniform_bounds}. Here the center point is chosen to be $c_1 = \ldots = c_K = 10$, the radius of the $K$-ball is set to $r = 2$, and $K$ varies between $1$ and $100$. Clearly, the lower bound becomes tighter for large $K$. In fact, it is not hard to show that 
\begin{equation}
  \frac{\varepsilon_K(c,r)}{K} \to 0 \quad \text{for} \quad K \to \infty,
\end{equation}
meaning the lower bound is asymptotically tight and coincides with the upper bound. Hence
\begin{equation}
  \lim_{K \to \infty} \mmse_{X|Y}(P_{XY}) = r^2.
  \label{eq:uniform_mmse_limit}
\end{equation}

While this result could have been obtained in a more straightforward manner, if allows for some interesting insights. The limit in \eqref{eq:uniform_mmse_limit} implies that, asymptotically, the MMSE estimator for the model in \eqref{eq:mulitplicative_channel} is a constant, namely $f^*(y) = \mu_X = c$. Interestingly, the aspect that the observations contain a vanishingly small amount of information is captured by the Gaussian approximation model, where $X_0$ and $Y_0$ are entirely independent. Nevertheless, the distribution of $Y$ is of importance since it contributes to the distance between the approximated and the true joint distribution. Hence, the proposed bounds capture the asymptotic independence of input and output, while using the Gaussian approximation to bound the impact of ignoring this dependence for finite $K$.

\begin{figure}[tb]
    \centering
    \includegraphics{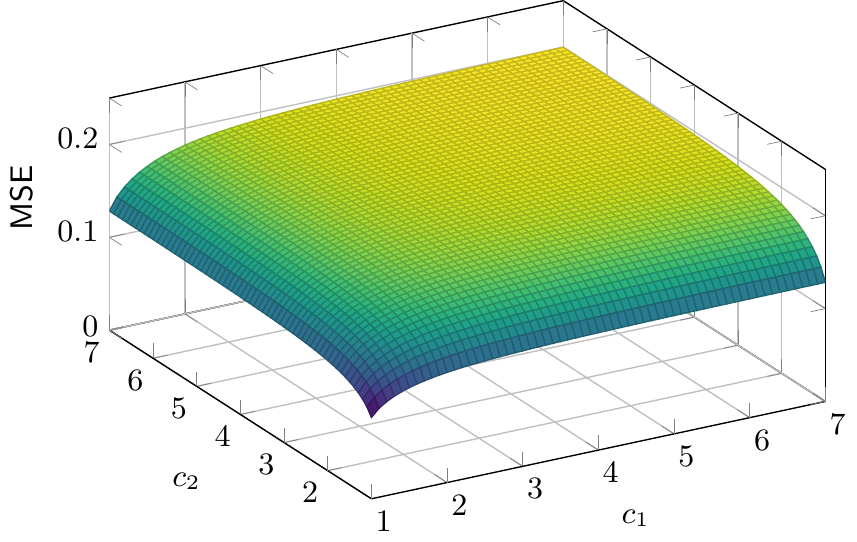}
    \caption{MMSE lower bound in \eqref{eq:uniform_lower_bound} with $K = 2$ and $r = 1$ for different center points $c = (c_1, c_2)$.} 
    \label{fig:uniform_bound_vs_c}
\end{figure}

The influence of the center point $c$ on the lower bound is illustrated in Fig.~\ref{fig:uniform_bound_vs_c} for the case $K = 2$ and $r = 1$. The bound is lower towards the axis, where small values of $X$ lead to small variances of $Y$, which in turn makes estimating $X$ from $Y$ easier. This effect becomes less and less pronounced as $c$ moves away from the origin, thus increasing the variance of $Y$. Asymptotically, for $c \to \infty$, the lower bound again approaches the upper bound ($\sigma_{X_0} = 1/4$), meaning that $Y$ becomes increasingly uninformative. 

\subsection{High and Low SNR Behavior}
Next, we show that the proposed bounds can also be used to study the high and low SNR behavior of the MMSE. We also formally show that our bounds perform better than the Cram\'er-Rao bound in the low SNR regime.

Consider the additive channel in \eqref{eq:AGN}, with $N \sim \mathcal{N}(0,\sigma_N^2 I)$.  
Then, using the bounding \eqref{eq:mmse_bounds_scalar}, we have that as $\sigma_N^2$ approaches zero (the high SNR regime)
\begin{equation}
    \mmse_{X|Y}(P_{XY})= \Theta(\sigma_N^2).
\end{equation}
It also interesting to note that this result holds even if $P_X$ is allowed to vary with $\sigma_N^2$ as long as the KL divergence is uniformly bonded (i.e., $ \sup_{ \sigma_N^2}\DKL{P_{X, \sigma_N^2}}{Q_{X_0}}<\infty$). 

We now study the low SNR behaviour of our bounds. As a consequence of this analysis, we will show that in this regime our bounds are tighter than the Cram\'er-Rao bound.
First, choose   $\Sigma_{N_0}=\sigma_N^2 I$ and $\Sigma_{X_0}=\sigma_0^2 I$.   Second, using the lower bound in \eqref{eq:mmse_bounds_scalar}, we have that  
\begin{equation}
    \lim_{\sigma_N^2 \to \infty } \, \mmse_{X|Y}(P_X) \ge  \omega_0 \!\left( \frac{\varepsilon}{K} \right)  \sigma_0^2 K, \label{eq:limitLowSNR}
\end{equation}
where $\sigma_0^2$ is arbitrary and 
\begin{align}
    \varepsilon=\DKL{P_{X}}{Q_{X_0}}= - h(X)+\frac{K}{2}\log(2\pi \sigma_0^2) +\frac{\tr( \Sigma_X)}{2 \sigma_0^2}.
\end{align}
Taking $\sigma_0^2 \to \infty$ on the right side of \eqref{eq:limitLowSNR} leads to
\begin{align}
     \lim_{\sigma_N^2 \to \infty } \, \mmse_{X|Y}(P_X) \ge  \frac{1}{2\pi e} e^{\frac{2}{K} h(X)}.
\end{align}
The above procedure can now be compared to the Cram\'er-Rao bound, which leads to the following limit:
\begin{align}
  \lim_{\sigma_N^2 \to \infty } \, \mmse_{X|Y}(P_X) &\ge  \lim_{\sigma_N^2 \to \infty}  \tr \left( \left(\frac{1}{\sigma_N^2}I +I(P_X) \right )^{-1} \right  ) \\
 &=\tr\left ( I^{-1}(P_X)   \right ) .
\end{align}
Next, invoking Stam's inequality \cite{raginsky2012concentration} we have that
\begin{align}
    \frac{1}{2\pi e} e^{\frac{2}{K} h(X)} \ge  \tr\left ( I^{-1}(P_X)   \right ).
\end{align}
The above  discussion shows that proposed bounds are tighter than the Cram\'er-Rao bound at the low SNR regime.

\section{Conclusions and Outlook}
\label{sec:conclusion}
This work has considered the problem of maximizing and minimizing the MMSE when estimating a random vector $X \in \mathbb{R}^K$ form a random vector $Y \in \mathbb{R}^M$, subject to the constraint that their joint distribution $P_{XY}$ lies in a KL divergence ball of radius $\varepsilon$ centered at a Gaussian reference distribution. It has been shown that both the maximum and the minimum are attained by jointly Gaussian distributions whose mean is identical to that of the reference distribution and whose covariance matrix can be determined by finding a scalar root of a monotonic function. The upper bound has been shown to provide a minimax robust MMSE estimator under distributional uncertainty of the KL divergence ball type. The lower bound has been shown to be potentially superior to the Bayesian Cram\'er-Rao lower bound and to be defined for a larger family of distributions.

We conclude the paper by mentioning a few interesting directions for future research:
\begin{itemize}
  \item Since the MMSE is closely related to other estimation and information measures, such as Fisher information and mutual information, bounds on the latter can be derived from the bounds on the former. For the AGN channel, this aspect has already been explored in \cite{DytsoFauss2019}. The generalized bounds presented here allow for an extension to non-Gaussian and non-additive channels.
  
  \item The results in this paper as well as those in \cite{DytsoFauss2019} only apply to continuous distributions. This is the case since distributions that contain point masses admit an infinite KL divergence w.r.t.\ a Gaussian distribution. This naturally raises the question whether a similar bound can be established for discrete distributions. In particular, the question which reference distribution and which divergence measure to use in this case is of both practical and theoretical interest. 
    
  \item In general, the question arises whether the approach used in this paper can be extended to different cost functions, divergence measures, and reference distributions. This is a particularly interesting topic in light of a recent work of ours, in which the Cram\'er--Rao bound was generalized to Bregman divergences \cite{dytso2020class}. A corresponding result based on the approach followed here might provide alternative bounds and additional insights.  
\end{itemize}


%


%
%

\ifCLASSOPTIONcaptionsoff
  \newpage
\fi



\bibliographystyle{IEEEtran}
\bibliography{bibliography}
\end{document}